\documentclass[12pt]{iopart}

\usepackage{iopams}
\usepackage{setstack}
\usepackage{amscd}
\usepackage{amsfonts}
\usepackage{amssymb}
\usepackage{graphicx}
\usepackage{hyperref}
\usepackage{bm}

\def\be{\begin{equation}}
\def\ee{\end{equation}}
\def\ba{\begin{eqnarray}}
\def\ea{\end{eqnarray}}

\begin{document}

\newcommand{\vp}{\varphi}
\newcommand{\rd}{{\rm d}}

\title[Quintessence:~A Review]{Quintessence:~A Review}

\author{Shinji Tsujikawa}

\address{Department of Physics, Faculty of Science, 
Tokyo University of Science, 
1-3, Kagurazaka, Shinjuku-ku, Tokyo 162-8601, Japan}

\ead{shinji@rs.kagu.tus.ac.jp}

\begin{abstract}

Quintessence is a canonical scalar field introduced to explain
the late-time cosmic acceleration.
The cosmological dynamics of quintessence is reviewed, 
paying particular attention to the evolution of the dark 
energy equation of state $w$. For the field potentials having 
tracking and thawing properties, the evolution of $w$ can be 
known analytically in terms of a few model parameters.
Using the analytic expression of $w$, we constrain quintessence 
models from the observations of supernovae type Ia, 
cosmic microwave background, and baryon acoustic 
oscillations. The tracking freezing models are hardly 
distinguishable from the $\Lambda$-Cold-Dark-Matter 
($\Lambda$CDM) model, 
whereas in thawing models the today's field equation of state is 
constrained to be $w_0<-0.7$ (95 \%\,CL).
We also derive an analytic formula for the growth rate of 
matter density perturbations in dynamical dark 
energy models, which allows a possibility to put further
bounds on $w$ from the measurement of redshift-space
distortions in the galaxy power spectrum.
Finally we review particle physics models of quintessence--such 
as those motivated by supersymmetric theories.
The field potentials of thawing models based on 
a pseudo-Nambu-Goldstone boson or on extended supergravity 
theories have a nice property that a tiny mass of quintessence 
can be protected against radiative corrections.

\end{abstract}

\maketitle

\section{Introduction}
\label{aba:sec1}

The observational discovery of the late-time cosmic acceleration from 
the Supernovae type Ia (SN Ia) opened up a new research area 
in modern cosmology \cite{Riess,Perlmutter}.
About 70\,\% of the energy density of the Universe today 
consists of an unknown component called dark energy.
This has been also confirmed by other observations-- such as 
Cosmic Microwave Background (CMB) \cite{WMAP1,Planck} 
and Baryon Acoustic Oscillations (BAO) \cite{BAO1}. 
The property of dark energy is characterized by 
the equation of state $w=P/\rho$, where $P$ is 
the pressure and $\rho$ is the energy density.
Dark energy has a negative pressure with $w$
less than $-1/3$.

One of the simplest candidates of dark energy is the cosmological 
constant $\Lambda$ with $w=-1$.
The cosmological constant can arise from a vacuum energy in 
particle physics, but its energy scale is enormously larger 
than the observed energy scale of dark energy \cite{Weinberg}.
There have been many attempts to construct de Sitter vacua in 
supersymmetric theories.
In string theory, for example, huge numbers of de Sitter vacua 
($\sim 10^{500}$) can be present after the so-called flux compactification 
of higher-dimensional manifolds \cite{KKLT}. 
We may live in a vacuum with a tiny vacuum energy, but it is 
generally difficult to justify the reason for living 
such a specific vacuum unless some anthropic principle 
is introduced. So far, it is fair to say that there is
no satisfactory scenario where the small energy 
scale of dark energy can be naturally explained by 
the vacuum energy related to particle physics.

If the cosmological constant problem is solved in a way that 
it vanishes completely, we need to find out an alternative 
mechanism to explain the origin of dark energy \cite{darkreview1,darkreview2}.  
Broadly speaking, we can classify dark energy models 
into two classes. The first one is based on a specific form 
of matter-- such as 
quintessence \cite{Fujii,Ratra,CSN,Ferreira,CLW,Caldwell,Zlatev}, 
k-essence \cite{kes1,kes2}, and the Chaplygin gas \cite{chap}. 
The second one is based on the modification of gravity 
at large distances (see Refs.~\cite{moreview} for reviews).
In both classes the dark energy equation of 
state dynamically changes in time, by which the models 
can be distinguished from the 
$\Lambda$CDM model.

Quintessence is described by a canonical scalar 
field $\phi$ minimally coupled to gravity.
Compared to other scalar-field models such as 
phantoms and k-essence, quintessence is the simplest 
scalar-field scenario without having theoretical problems 
such as the appearance of ghosts and Laplacian instabilities. 
A slowly varying field along a potential $V(\phi)$ 
can lead to the acceleration of the Universe. 
This mechanism is similar to slow-roll inflation 
in the early Universe, but the difference is that 
non-relativistic matter (dark matter and baryons) 
cannot be ignored to discuss the dynamics of dark energy correctly. 
Moreover, the energy scale of the quintessence potential 
needs to be of the order of $\rho_{\rm DE} \approx 10^{-47}$~GeV$^4$ 
today, which is much smaller than that of the inflaton potential.

The dynamics of quintessence in the presence of non-relativistic
matter has been studied in detail for many different 
potentials \cite{CLW,Caldwell,Zlatev,Macorra,Nunes,Cora,CLinder,Linder06}. 
Depending on the evolution of $w$, we can broadly 
classify quintessence models into two classes \cite{CLinder}:
(i) thawing models and (ii) freezing models.
In the first class, the field is nearly frozen by a Hubble 
friction during the early cosmological epoch and it starts to 
evolve once the field mass drops below the Hubble expansion rate.
In the second class, the evolution of the field gradually slows 
down because the potential tends to be shallow at late times.
For the inverse power-law potential $V(\phi)=M^{4+p}\phi^{-p}$ 
($p>0$), there is so-called a tracker solution \cite{SWZ} 
along which $w$ is nearly constant during the matter era 
and $w$ starts to decrease after that.
This case belongs to a subclass of freezing models.
For thawing and tracker models there exist convenient 
analytic formulas of $w$ \cite{SchSen,Dutta,Chiba09,Chiba10} 
employed to test the models with the data of 
distance measurements of SN Ia, CMB, and BAO.

The redshift-space distortions (RSD) appearing in clustering pattern of 
galaxies \cite{Kaiser,Tegmark06} can provide additional constraints on 
the growth rate of matter perturbations $\delta_m$.
Since the evolution of $\delta_m$ is different depending 
on the field equation of state \cite{Wang98,Linder,Gong}, 
it is possible to place bounds on $w$
from the data of RSD. 
In fact there exist analytic formulas of $\delta_m$ and its 
growth rate \cite{Jailson}, which can be used to constrain
quintessence models.

In order to realize the cosmic acceleration today, the mass 
$m_{\phi}$ of quintessence (defined by $m_{\phi}^2=d^2 V (\phi)/d\phi^2$)
needs to be extremely small, i.e., $|m_{\phi}| \lesssim H_0 \approx 10^{-33}$\,eV, 
where $H_0$ is the today's Hubble parameter.
In general there is a difficulty to reconcile such a ultra light mass
with the energy scales appearing in particle 
physics \cite{Carrollqui}.
Moreover, in the absence of some symmetry, the radiative corrections 
may disrupt the flatness of the quintessence potentials required 
for the cosmic acceleration \cite{Kolda}.
However, it is not entirely hopeless to construct viable quintessence 
models in the framework of particle 
physics \cite{Frieman,Brax,Nomura,Choi,Kim,Hall,CNR,Townsend,Panda}.

In this article, we review several cosmological aspects of quintessence--
including its cosmological dynamics, analytic solutions of $w$, 
observational constraints, and particle physics models.
The review is organized as follows.
In Sec.\,\ref{aba:sec2} we present the field equations of motion 
for general quintessence potentials and then proceed to 
the analysis of fixed points for exponential potentials.
In Sec.\,\ref{aba:sec3} we classify quintessence potentials 
into two classes depending on the evolution of $w$ and 
then derive analytic solutions of $w$. 
These solutions are employed to put observational 
bounds on quintessence at the background level.
In Sec.\,\ref{aba:sec4} we derive analytic formulas for the growth 
rate of matter density perturbations and discuss constraints
on some of quintessence models from the recent data of RSD.
In Sec.\,\ref{aba:sec5} we review theoretical models of 
quintessence based on supersymmetric theories.
Sec.~\ref{aba:sec6} is devoted to conclusions.

\section{Dynamical equations of motion
and exponential potentials}
\label{aba:sec2}

Let us consider quintessence in the presence of non-relativistic
matter described by a barotropic perfect fluid.
The total action is given by 
\be
S=\int d^4 x \sqrt{-g}
\left[ \frac{1}{2}M_{\rm pl}^2 R-\frac12 g^{\mu \nu} 
\partial_{\mu} \phi \partial_{\nu} \phi-V(\phi) 
\right]+S_m\,,
\label{quinac}
\ee
where $g$ is the determinant of the metric $g_{\mu \nu}$, 
$M_{\rm pl}$ is the reduced Planck mass, $R$ is the 
Ricci scalar, $S_m$ is the matter action.
We assume that non-relativistic matter does not 
have a direct coupling to the quintessence field $\phi$.

We study the dynamics of quintessence on the flat 
Friedmann-Lema\^{i}tre-Robertson-Walker (FLRW)
background with the line element 
$ds^2=-dt^2+a^2(t) d{\bm x}^2$, where 
$a(t)$ is the scale factor with cosmic time $t$.
The pressure and the energy density of quintessence
are given, respectively, by 
$P_{\phi}=\dot{\phi}^2/2-V(\phi)$ and 
$\rho_{\phi}=\dot{\phi}^2/2+V(\phi)$, where a dot 
represents a derivative with respect to $t$.
The dark energy equation of state is 
\be
w \equiv \frac{P_{\phi}}{\rho_{\phi}}=
\frac{\dot{\phi}^2/2-V(\phi)}{\dot{\phi}^2/2+V(\phi)}\,.
\label{wphiquin}
\ee
The scalar field satisfies the continuity equation 
$\dot{\rho_{\phi}}+3H(\rho_{\phi}+P_{\phi})=0$, i.e., 
\be
\ddot{\phi}+3H \dot{\phi}+V_{,\phi}=0\,,
\label{stanba2}
\ee
where $H \equiv \dot{a}/a$ and 
$V_{,\phi} \equiv dV/d\phi$.
For a matter fluid with the energy density $\rho_m$ and 
the equation of state $w_m$, the equations of motion 
following from the action (\ref{quinac}) are
\ba
3M_{\rm pl}^2 H^2 &=& \dot{\phi}^2/2+V(\phi)+\rho_m 
\,,\label{quin1} \\
2M_{\rm pl}^2\dot{H} &=&
-\left[ \dot{\phi}^2+(1+w_m)\rho_m \right]\,.
\label{quin2}
\ea

In order to deal with the cosmological dynamics of
this system, it is convenient to introduce the following 
dimensionless variables \cite{CLW}
\be
x \equiv \frac{\dot{\phi}}{\sqrt{6}M_{\rm pl}H}\,,\qquad
y \equiv \frac{\sqrt{V(\phi)}}{\sqrt{3}M_{\rm pl}H}\,.
\ee
The field density parameter $\Omega_{\phi} \equiv \rho_{\phi}/
(3M_{\rm pl}^2 H^2)$ can be expressed as 
$\Omega_{\phi}=x^2+y^2$.
From Eq.~(\ref{quin1}) the matter density parameter
$\Omega_m \equiv \rho_m/(3M_{\rm pl}^2 H^2)$ 
satisfies $\Omega_m=1-\Omega_{\phi}$.
The field equation of state (\ref{wphiquin}) reads
$w=(x^2-y^2)/(x^2+y^2)$.
We also define the effective equation of state 
$w_{\rm eff} \equiv -1-2\dot{H}/(3H^2)$, where 
$\dot{H}/H^2$ can be evaluated from 
Eq.~(\ref{quin2}) as
\be
\frac{\dot{H}}{H^2}=-3x^2-\frac32 (1+w_m)
(1-x^2-y^2)\,.
\label{dothquin}
\ee
Taking the derivatives of $x$ and $y$ with respect to 
$N \equiv \ln a$ and using Eqs.~(\ref{stanba2}) and 
(\ref{dothquin}), it follows that 
\ba
\frac{dx}{dN} &=& -3x+\frac{\sqrt{6}}{2}\lambda y^2
+\frac32 x \left[ (1-w_m)x^2+(1+w_m)(1-y^2) \right], 
\label{dxN} \\
\frac{dy}{dN} &=& -\frac{\sqrt{6}}{2}\lambda xy
+\frac32 y  \left[ (1-w_m)x^2+(1+w_m)(1-y^2) \right]\,,
\label{dyN} 
\ea
where $\lambda$ is defined by 
$\lambda \equiv -M_{\rm pl}V_{,\phi}/V$.

The models with constant $\lambda$ corresponds to 
the exponential potential \cite{Ferreira,CLW,Lucchin}
\be
V(\phi)=V_0 e^{-\lambda \phi/M_{\rm pl}}\,,
\label{expo}
\ee
in which case Eqs.~(\ref{dxN}) and (\ref{dyN}) are closed.
The fixed points of this system can be derived by 
setting $dx/dN=0$ and $dy/dN=0$ \cite{CLW}:
\begin{itemize}
\item (a)~$(x, y)=(0,0)$,\quad{}$\Omega_{\phi}=0$,\quad{}
$w_{{\rm eff}}=w_{m}$,
\quad $w$ is undetermined. 
\item (b)~$(x, y)=(\pm 1,0)$,\quad{}
$\Omega_{\phi}=1$,\quad{}$w_{{\rm eff}}=w=1$.
\item (c)~$(x, y)=(\lambda/\sqrt{6},[1-\lambda^{2}/6]^{1/2})$,
\quad{}$\Omega_{\phi}=1$,\quad
$w_{{\rm eff}}=w=-1+\lambda^{2}/3$.
\item (d)~$(x, y)=(\sqrt{3/2}(1+w_{m})/\lambda,
[3(1-w_{m}^{2})/2\lambda^{2}]^{1/2})$, \\
$\Omega_{\phi}=3(1+w_{m})/\lambda^{2}$,\quad{}
$w_{{\rm eff}}=w=w_{m}$. \label{scalingo}
\end{itemize}
If we consider non-relativistic matter ($w_m=0$), 
the matter-dominated epoch 
($w_{\rm eff} \simeq 0, \Omega_{\phi} \ll 1$) 
can be realized either by (a) or (d).
The point (d) is the so-called scaling solution \cite{Ferreira,CLW}, 
along which the ratio $\Omega_m/\Omega_{\phi}\,(\neq 0)$ 
remains constant. 
In order to realize the matter-dominated epoch by the scaling solution, 
we require the condition $\lambda^2 \gg 1$.
On the other hand, under the condition $\lambda^2<2$, 
the epoch of cosmic acceleration ($w_{\rm eff}<-1/3$) can 
be realized by the point (c).
This shows that the transition from (d) to (c) is not 
possible, but for $\lambda^2<2$ the system can 
evolve from (a) to (c).
The radiation-dominated epoch corresponds to the 
fixed point (a) with $w_{\rm eff}=w_m=1/3$.

In order to study the stabilities of the fixed points 
$(x,y)=(x_c,y_c)$, we consider linear perturbations 
$\delta x$ and $\delta y$
about them. Then the perturbations 
satisfy the following differential equations
\be
\hspace{-1cm}\frac{d}{d N}\left(\begin{array}{c}
\delta x \\
\delta y 
\end{array}\right)={\cal M}\left(\begin{array}{c}
\delta x \\
\delta y 
\end{array}\right)\,,\qquad
{\cal M}=\left(\begin{array}{cc}
\partial f_1/\partial x & \partial f_1/\partial y \\
\partial f_2/\partial x & \partial f_2/\partial y
\end{array}\right)_{x=x_c,\,y=y_c}\,,
\label{uvdif}
\ee
where $f_1(x,y)$ and $f_2(x,y)$ are the r.h.s. of 
Eqs.~(\ref{dxN}) and (\ref{dyN}) respectively.
If both the eigenvalues $\mu_1$ and $\mu_2$
of the matrix ${\cal M}$ are negative, 
the corresponding fixed point is stable. 
If either $\mu_1$ or $\mu_2$ is negative, 
the point corresponds to a saddle.
If both $\mu_1$ and $\mu_2$ are positive, 
the fixed point is unstable.
For complex values of $\mu_1$ and $\mu_2$ with 
negative real parts, the fixed point is called 
a stable spiral.

The eigenvalues of the point (c) are 
$\mu_1=(\lambda^2-6)/2$ and 
$\mu_2=\lambda^2-3(1+w_m)$ \cite{CLW}, so that 
it is stable under the condition 
$\lambda^2<3(1+w_m)$ for $0 \le w_m \le 1$.
The condition for cosmic acceleration corresponds
to $\lambda^2<2$, in which case the point (c) 
is stable. The eigenvalues of the point (a) 
are $\mu_1=-(3/2)(1-w_m)$, $\mu_2=(3/2)(1+w_m)$, 
so that it is a saddle for $0 \le w_m \le 1$.
This means that, for $\lambda^2<2$, the solution 
eventually exits the point (a) to approach 
the attractor point (c).

The dark energy equation of state $w$ for the point (a)
is undetermined, but in the realistic Universe,
$x$ and $y$ are not exactly 0.
The early evolution of $w$ depends on the initial 
conditions of $x$ and $y$.
If $x^2 \gg y^2$ and $x^2 \ll y^2$, we have 
$w \simeq 1$ and $w \simeq -1$ respectively.
Finally the solution approaches the constant value 
$w=-1+\lambda^2/3$.
Since $w$ dynamically changes in this way, the quintessence 
model with the exponential potential is observationally 
distinguishable from the $\Lambda$CDM model. 

For quintessence models in which $\lambda$ is not 
constant, Eqs.~(\ref{dxN}) and (\ref{dyN}) are not closed.
In such cases the situation is more involved, but 
it is possible to derive analytic solutions of $w$
by classifying quintessence potentials according to 
the evolution of $w$.
In the next section we shall address this issue.

\section{Classification of quintessence models and 
observational constraints}
\label{aba:sec3}

In order to study the evolution of $w$ for 
the models with varying $\lambda$, we derive
the differential equations for $w$ and $\Omega_{\phi}$. 
Using Eqs.~(\ref{dothquin})-(\ref{dyN}), we obtain 
\ba
w'  &=& (w-1) [ 3(1+w)
-\lambda \sqrt{3(1+w)\Omega_{\phi}} ]\,,
\label{quinw}\\
\Omega_{\phi}' &=& -3(w-w_m) \Omega_{\phi}
(1-\Omega_{\phi})\,,
\label{quinOme}
\ea
where a prime represents a derivative with respect to 
$N=\ln a$.
Introducing the quantity $\Gamma \equiv 
VV_{,\phi \phi}/V_{,\phi}^2$, 
the parameter $\lambda$ obeys
\be
\lambda' = -\sqrt{3(1+w) \Omega_{\phi}}\,
(\Gamma-1) \lambda^2\,.
\label{quinlam}
\ee
The evolution of $w$ is different depending on the quintessence potentials 
and the initial conditions.
In what follows we discuss three qualitatively different cases: 
(i) tracking freezing models, (iii) scaling freezing models, and 
(iii) thawing models.
In freezing models the potential tends to be shallow at late times, 
which results in the decrease of $w$.
In thawing models the mass of the field becomes smaller than 
$H$ only recently, so that the deviation of $w$ from $-1$ occurs
at late times.

\subsection{Tracking freezing models}

For the field density parameter satisfying the relation
\be
\Omega_{\phi}=3(1+w)/\lambda^2\,,
\label{Ometra}
\ee
$w$ is constant from Eq.~(\ref{quinw}).
If $w=w_m$, then $\Omega_{\phi}$ is constant from 
Eq.~(\ref{quinOme}) and hence $\lambda$ is constant.
This corresponds to the scaling solution (d) discussed 
in Sec.\,\ref{aba:sec2}.  
Since the scaling solution is stable for 
$\lambda^2>3(1+w_m)$ \cite{CLW}, 
it does not exit to the fixed point (c).
If $\lambda$ decreases in time, the system can 
enter the epoch of cosmic acceleration.
{}From Eq.~(\ref{quinlam}) this condition translates into
\be
\Gamma>1\,.
\label{Gamcon}
\ee
The solution (\ref{Ometra}) satisfying the condition (\ref{Gamcon})
is called a tracker \cite{SWZ}, 
along which $\Omega_{\phi}$ increases and hence $w<w_{m}$.
The tracker corresponds to a common evolutionary trajectory that 
attracts the solutions with different initial conditions.
{}From Eq.~(\ref{Ometra}) we have the relation 
$\Omega_{\phi}'/\Omega_{\phi}=-2\lambda'/\lambda$.
Using Eqs.~(\ref{quinOme}) and (\ref{quinlam})
under the condition $\Omega_{\phi} \ll 1$, the constant 
equation of state along the tracker is \cite{SWZ}
\be
w=w_{(0)} \equiv
\frac{w_m-2(\Gamma-1)}{2\Gamma-1}\,.
\label{w0tra}
\ee
For example, let us consider the inverse 
power-law potential \cite{Ferreira}
\be
V(\phi)=M^{4+p} \phi^{-p}\,,
\label{inversepo}
\ee
where $M$ and $p~(>0)$ are constants.
Since in this case $\Gamma=1+1/p>1$, the tracking 
condition (\ref{Gamcon}) is satisfied.
The constant equation of state (\ref{w0tra})
is $w_{(0)}=(p\,w_m-2)/(p+2)$ and hence 
$w_{(0)}=-2/(p+2)$ during the matter era.
With the growth of $\Omega_{\phi}$, $w$ 
starts to decrease from $w_{(0)}$.
Hence the tracker belongs to the class of 
freezing models.

\begin{figure}[htbp]
\begin{center}
\includegraphics[width=80mm]{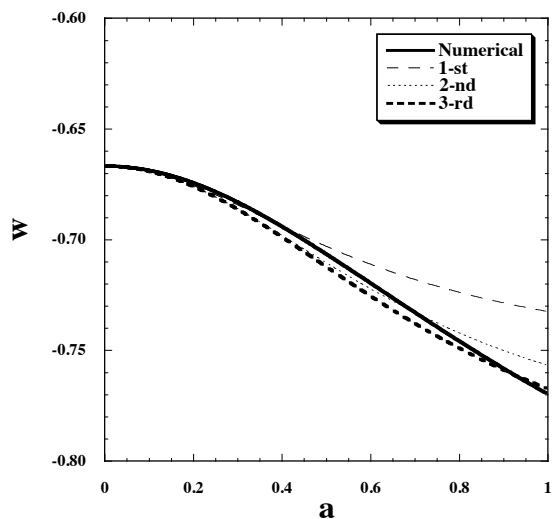}
\end{center}
\caption{The field equation of state $w$ versus $a$ 
for the tracker solution. 
This case corresponds to the inverse power-law 
potential $V(\phi)=M^5 \phi^{-1}$. 
The solid curve is derived by solving 
Eqs.~(\ref{quinw})-(\ref{quinlam}) numerically, 
whereas other curves show the 1-st, 2-nd, 3-rd 
order analytic solutions (\ref{wtracker}).}
\label{fig1}
\end{figure}

The analytic solution (\ref{w0tra}) is valid in the regime
$\Omega_{\phi} \ll 1$. We take into account the 
variation of $w$ by dealing with $\Omega_{\phi}$
as a perturbation to the 0-th order solution (\ref{w0tra}).
From Eqs.~(\ref{quinw}) and (\ref{quinlam}), 
the perturbation $\delta w$ around $w_{(0)}$ 
obeys \cite{Chiba10}
\be
\hspace{-1cm}
a^2 \frac{d^2 \delta w}{d a^2}+\frac{5-6w_{(0)}}
{2}a \frac{d \delta w}{da}+\frac92 (1-w_{(0)})
\delta w-\frac92 w_{(0)} (1-w_{(0)}^2)
\Omega_{\phi}(a)=0\,,
\label{deltaweq}
\ee
where we assumed that $\Gamma$ is nearly constant.
For $\Omega_{\phi} (a)$ we use the 0-th order solution
\be
\Omega_{\phi}(a)=\frac{\Omega_{\phi 0}\,a^{-3w_{(0)}}}
{\Omega_{\phi 0}\, a^{-3w_{(0)}}+
1-\Omega_{\phi 0}}\,,
\label{Omephia}
\ee
where $\Omega_{\phi 0}$ is the today's value of $\Omega_{\phi}$.
Substituting Eq.~(\ref{Omephia}) into Eq.~(\ref{deltaweq}), we obtain 
the following integrated solution \cite{Chiba10}
\be
w(a)=w_{(0)}+\sum_{n=1}^{\infty}
\frac{(-1)^{n-1}w_{(0)}(1-w_{(0)}^2)}
{1-(n+1)w_{(0)}+2n(n+1)w_{(0)}^2}
\left( \frac{\Omega_{\phi}(a)}{1-\Omega_{\phi}(a)}
\right)^n\,.
\label{wtracker}
\ee

In Fig.~\ref{fig1} we plot the evolution of $w$ derived 
by the analytic solution (\ref{wtracker}) for the 
inverse power-law potential $V(\phi)=M^5\phi^{-1}$.
Each curve corresponds to the 1-st, 2-nd, 3-rd order solution, 
whereas the solid curve is derived by solving 
Eqs.~(\ref{quinw})-(\ref{quinlam}) numerically.
We find that the analytic solution up to 3-rd 
order shows good agreement with the full numerical result. 
The analytic expression of $w$ is parametrized 
by two parameters $w_{(0)}$ and $\Omega_{\phi 0}$ alone.

The observational constraints on the tracker model have 
been carried out in Refs.~\cite{Chiba10,PWang,CDT}.
In addition to the SN Ia data, the distance measurements of 
the CMB and BAO peaks provide the information of the 
background expansion history from the recombination 
epoch to today.
{}From the joint data analysis of Union 2.1 \cite{Suzuki}, 
WMAP7 \cite{WMAP7}, and BAO 
(SDSS7 \cite{SDSS7} and BOSS \cite{BOSS}), 
the tracker equation of state during the matter era 
is constrained to be $w_{(0)}<-0.964$ (95\% CL) under the prior
$w_{(0)}>-1$ \cite{CDT}. For the potential (\ref{inversepo})
this bound translates into $p<0.075$. 
In Ref.~\cite{CDT} it was found that the best-fit corresponds 
to $w_{(0)}=-1$, i.e., the $\Lambda$CDM.
If we do not put the prior $w_{(0)}>-1$, the best-fit model 
parameters are found to be $w_{(0)}=-1.097$ and 
$\Omega_{\phi 0}=0.717$.
With the BOSS BAO data  \cite{BOSS} the phantom equation of 
state ($w_{(0)}<-1$) 
is particularly favored, but this is not the regime of quintessence.

\subsection{Scaling freezing models}

The scaling solution \cite{Ferreira,CLW} can be regarded as 
a special case of a tracker along which 
$\Omega_{\phi}=3(1+w)/\lambda^2$ is constant. 
During the matter era, $w=w_m=0$ and hence
$\Omega_{\phi}=3/\lambda^2$. 
Since $\lambda$ is constant,
$\Gamma=1$ from Eq.~(\ref{quinlam}).
This case corresponds to the exponential 
potential (\ref{expo}), but the system does not enter 
the phase of cosmic acceleration because the field equation 
of state is the same as that of the background fluid.

This problem can be alleviated by considering 
the double exponential potential \cite{BCN}
\be
V(\phi)=V_1 e^{-\lambda_1 \phi/M_{\rm pl}}+
V_2 e^{-\lambda_2 \phi/M_{\rm pl}}\,,
\label{doublepo}
\ee
where $\lambda_i$ and $V_i$ ($i=1, 2$) are 
constants (see Refs.~\cite{SahniWang,Albrecht,Stewart} 
for related potentials).
For the parameters satisfying the conditions 
$\lambda_1 \gg 1$ and $\lambda_2 \lesssim 1$, 
the solution first enters the scaling regime 
characterized by $\Omega_{\phi}=3(1+w_m)/\lambda_1^2$. 
During the radiation era ($w_m=1/3$) the constraint 
coming from the big bang nucleosynthesis gives the bound
$\Omega_{\phi}<0.045$ (95\,\% CL) \cite{Bean},
which translates into the condition $\lambda_1>9.4$.
The scaling matter era ($\Omega_{\phi}=3/\lambda_1^2$, $w=0$)
is followed by the epoch of cosmic acceleration
driven by another exponential potential 
$V_2 e^{-\lambda_2 \phi/M_{\rm pl}}$.
In this case the solution finally approaches the fixed point 
(c) discussed in Sec.\,\ref{aba:sec2}.

\begin{figure}[htbp]
\begin{center}
\includegraphics[width=80mm]{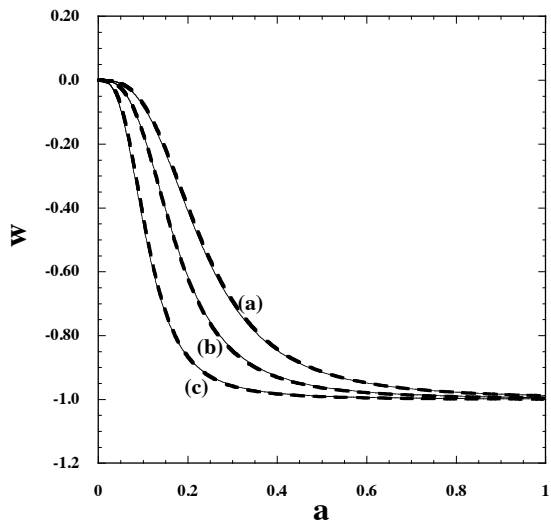}
\end{center}
\caption{The field equation of state $w$ versus
$a$ for the potential (\ref{doublepo}) with 
(a) $\lambda_1=10$, $\lambda_2=0$, 
(b) $\lambda_1=15$, $\lambda_2=0$, and 
(c) $\lambda_1=30$, $\lambda_2=0$.
The solid curves are the numerically integrated solutions, 
whereas the dashed curves show the results derived from 
the parametrization (\ref{LHpara}) with $w_p=0$ and $w_f=-1$.
Each dashed curve corresponds to 
(a) $a_t=0.23$, $\tau=0.33$, 
(b) $a_t=0.17$, $\tau=0.33$, and 
(c) $a_t=0.11$, $\tau=0.32$.}
\label{fig2}
\end{figure}

The onset of the transition from the scaling matter era to 
the epoch of cosmic acceleration depends on 
the parameters $\lambda_1$, $\lambda_2$, and $V_2/V_1$.
The transition redshift is not very sensitive to the choice of 
$V_2/V_1$, so we can set $V_2=V_1$ without 
loss of generality.
In Fig.~\ref{fig2} we show the numerical evolution of $w$
for $\lambda_2=0$ with three different values of $\lambda_1$.
For larger $\lambda_1$ the transition to $w=-1$ occurs earlier.

The above variation of $w$ can be accommodated
by using the parametrization \cite{Linder05} 
\be
w(a)=w_f+\frac{w_p-w_f}{1+(a/a_t)^{1/\tau}}\,,
\label{LHpara}
\ee
where $w_p$ and $w_f$ are asymptotic values of $w$
in the past and future respectively, $a_t$ is the scale
factor at the transition, and $\tau$ describes the transition width 
(see Refs.~\cite{Bassett} for early related works).
The scaling solution during the matter-dominated epoch
corresponds to $w_p=0$.
For $\lambda_2=0$ we have $w_f=-1$, in which case 
Eq.~(\ref{LHpara}) reduces to 
$w(a)=-1+[1+(a/a_t)^{1/\tau}]^{-1}$.
As we see in Fig.~\ref{fig2}, the parametrization 
(\ref{LHpara}) fits the numerical solutions of $w$
very well for appropriate choices of $a_t$ and $\tau$.
For the models with $\lambda_2=0$ the transition width 
is around $\tau \approx 0.33$, 
while $a_t$ depends on $\lambda_1$.

In Ref.~\cite{CDT} the joint data analysis of Union 2.1, WMAP7, 
and BAO (SDSS7 and BOSS) was carried out by fixing $\tau=0.33$. 
The transition redshift was found to be $a_t<0.23$ (95\,\%\,CL).
The case (a) shown in Fig.~\ref{fig2} is the marginal one 
where the model is within the $2\sigma$ observational contour. 
This shows that $w$ needs to approach $-1$ in the early 
cosmological epoch. 
For $\lambda_2>0$ the likelihood analysis was also performed
in Ref.~\cite{CDT} by numerically solving the field equations of motion 
with suitable initial conditions.
The model parameters are constrained to be $\lambda_1>11.7$, 
$\lambda_2<0.539$, and $0.256<\Omega_{m0}<0.279$ (95\,\%\,CL).
The models with $\lambda_2 \gtrsim 0.5$ are disfavored because
the deviation of $w$ from $-1$ tends to be significant.

In k-essence models where the Lagrangian $P$
depends on the field $\phi$ and the kinetic energy 
$X=-g^{\mu \nu} \partial_{\mu} \phi \partial_{\nu} \phi/2$ \cite{kes1,kes2}, 
the condition for the existence of scaling solutions restricts
the Lagrangian to the form 
$P(\phi,X)=Xg(Xe^{\lambda \phi/M_{\rm pl}})$ \cite{Piazza,Waga}, 
where $g$ is an arbitrary function in terms of $Y \equiv Xe^{\lambda \phi/M_{\rm pl}}$.
The quintessence with the exponential potential ($P=X-c e^{-\lambda \phi/M_{\rm pl}}$) 
corresponds to the choice $g(Y)=1-c/Y$, whereas the choice $g(Y)=-1+cY$
gives rise to the dilatonic ghost condensate model 
$P=-X+ce^{\lambda \phi/M_{\rm pl}}X^2$ \cite{Piazza}.
For the multi-field scaling Lagrangian given by 
\be
P(\phi_i, X_i)=\sum_{i=1}^{n} X_i\, g(X_i e^{\lambda_i \phi_i/
M_{\rm pl}})\,,
\ee
it was shown \cite{Tsuji06} that a phenomenon called 
{\it assisted inflation} \cite{assisted} occurs with the effective slope 
$\lambda_{\rm eff}=(\sum_{i=1}^n 1/\lambda_i^2)^{-1/2}$, 
irrespective of the form of $g$. 
In the presence of multiple fields, the scaling matter era can 
be followed by the epoch of cosmic acceleration even if the 
individual field is unable to lead to the accelerated expansion.
In Refs.~\cite{Junko,Blais} the cosmological dynamics of 
assisted dark energy was studied in detail.

\subsection{Thawing models}
\label{thawingsec}

In thawing models the field is nearly frozen by the Hubble
friction in the early cosmological epoch. 
In this regime one has $w \simeq -1$, which corresponds 
to one of the fixed points of (\ref{quinw}).
The representative model of this class is characterized by 
the potential of the pseudo-Nambu-Goldstone boson 
(PNGB) \cite{Frieman}:
\be
V(\phi)=\mu^4 \left[ 1+\cos (\phi/f_a) \right]\,,
\label{pngbpo}
\ee
where $\mu$ and $f_a$ are constants having a dimension of mass.

Let us consider the case in which the field initially exists
around $\phi=\phi_i$ and then it starts to evolve after the field mass
drops below $H$. We expand the potential $V(\phi)$ around 
$\phi=\phi_i$ up to second order, as 
$V(\phi)=\sum_{n=0}^2 V^{(n)}(\phi_i)\,(\phi-\phi_i)^n/n!$.
Using the approximation $P_{\phi} \simeq -\rho_{\phi} \simeq -V(\phi_i)$
and redefining the field $u=(\phi-\phi_i)a^{3/2}$, 
Eq.~(\ref{stanba2}) reads \cite{Dutta,Chiba09}
\be
\ddot{u}-\omega^2u \simeq
-a^{3/2} V_{,\phi} (\phi_i)\,,\quad
\quad
\omega=\left[ \frac{3}{4}
\frac{V(\phi_i)}{M_{\rm pl}^2}
-V_{,\phi \phi} (\phi_i) \right]^{1/2}\,,
\label{ddotu}
\ee
where we assumed 
$3V(\phi_i)/(4M_{\rm pl}^2)>V_{,\phi \phi}(\phi_i)$.
Provided that $|w+1| \ll 1$ the evolution of the scale factor
can be approximated as that of the $\Lambda$CDM model, 
in which case
$\dot{a}^2=H_0^2 a^2 [\Omega_{\phi 0}
+(1-\Omega_{\phi 0})a^{-3}]$ from Eq.~(\ref{quin1}). 
Integration of this equation gives
\be
a(t)=\left( \frac{1-\Omega_{\phi 0}}{\Omega_{\phi 0}}
\right)^{1/3} \sinh^{2/3} (t/t_{\Lambda})\,,
\qquad 
t_{\Lambda}=\frac{2M_{\rm pl}}{\sqrt{3V(\phi_i)}}\,.
\label{aap}
\ee

Substituting Eq.~(\ref{aap}) into Eq.~(\ref{ddotu}), we obtain 
the following solution 
\be
u(t)=A \sinh (\omega t)+B \cosh (\omega t)
+\sqrt{\frac{1-\Omega_{\phi 0}}{\Omega_{\phi 0}}}
\frac{V_{,\phi} (\phi_i)\,t_{\Lambda}^2}{\omega^2 t_{\Lambda}^2-1}
\sinh (t/t_{\Lambda})\,,
\ee
which is valid for $\omega t_{\Lambda} \neq 1$
(i.e., $V_{,\phi \phi}(\phi_i) \neq 0$).
The integration constants $A$ and $B$ are determined 
by the initial conditions $\phi(0)=\phi_i$ 
and $\dot{\phi}(0)=0$.
Then, the solution is
\be
\phi(t)=\phi_i+\frac{V_{,\phi} (\phi_i)}{V_{,\phi \phi} (\phi_i)}
\left[ \frac{\sinh (\omega t)}{\omega t_{\Lambda} 
\sinh (t/t_{\Lambda})}-1 
\right]\,.
\ee
Since $w+1 \simeq \dot{\phi}^2(t)/V(\phi_i)$ under
the approximation $\rho_{\phi} \simeq V(\phi_i)$, 
it follows that 
\be
\hspace{-2cm}
w+1 \simeq \frac34 \left[ \frac{V_{,\phi}(\phi_i)}
{\omega t_{\Lambda}V_{,\phi \phi} (\phi_i)} \right]^2
\left[ \frac{\omega t_{\Lambda} \cosh (\omega t) \sinh (t/t_{\Lambda})
-\sinh (\omega t) \cosh (t/t_{\Lambda})}{\sinh^2 (t/t_{\Lambda})}
\right]^2\,.
\label{wtex}
\ee
The field equation of state can be written as a function of $a$
by using the value $w_0$ today ($t=t_0$, $a=1$).
Introducing the dimensionless variables
\be
\hspace{-1cm}
K \equiv  \omega t_{\Lambda}=\sqrt{1-\frac43 \frac{M_{\rm pl}^2 
V_{,\phi \phi} (\phi_i)}{V(\phi_i)}}\,,\qquad
F(a) \equiv \sqrt{1+[(\Omega_{\phi 0})^{-1}-1]a^{-3}}\,,
\label{KFdef}
\ee
we obtain the relation $\omega t=K \sinh^{-1} 
\sqrt{a^3\,\Omega_{\phi 0}/(1-\Omega_{\phi 0})}$ from Eq.~(\ref{aap}).
Then Eq.~(\ref{wtex}) reads \cite{Dutta,Chiba09}
\be
w(a) =-1+(1+w_0) a^{3(K-1)}{\cal F}(a)\,,
\label{waap}
\ee
where 
\be
{\cal F}(a)= \left[
\frac{(K-F(a))(F(a)+1)^K+(K+F(a))(F(a)-1)^K}
{(K-\Omega_{\phi 0}^{-1/2})(\Omega_{\phi 0}^{-1/2}+1)^K
+(K+\Omega_{\phi 0}^{-1/2})(\Omega_{\phi 0}^{-1/2}-1)^K}
\right]^2\,.
\ee
\begin{figure}[htbp]
\begin{center}
\includegraphics[width=80mm]{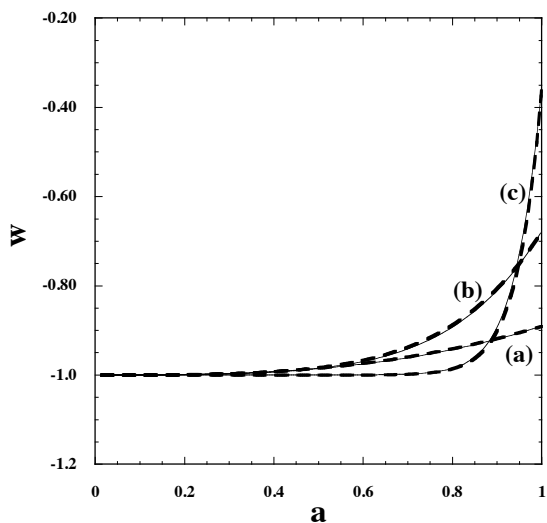}
\end{center}
\caption{The field equation of state $w$ versus
$a$ for the potential (\ref{pngbpo}) with 
(a) $f_a/M_{\rm pl}=0.5$, $\phi_i/f_a=0.5$ ($K=1.9$),
(b) $f_a/M_{\rm pl}=0.3$, $\phi_i/f_a=0.25$ ($K=2.9$), and 
(c) $f_a/M_{\rm pl}=0.1$, $\phi_i/f_a=7.6 \times 10^{-4}$ 
($K=8.2$). The solid curves correspond to numerically integrated 
solutions, whereas the bald dashed curves show the results 
derived from the analytic solution (\ref{waap}) 
with $\Omega_{\phi 0}=0.73$.}
\label{fig3}
\end{figure}

The field equation of state (\ref{waap}) is expressed 
in terms of the three parameters $w_0$, $\Omega_{\phi 0}$, 
and $K$. The quantity $K$ is related to the field mass 
squared $m_{\phi}^2=V_{,\phi \phi}$.
For the potential (\ref{pngbpo}) we have 
$K>1$ for $0<\phi_i/f_a<\pi/2$ and $K<1$ for
$\pi/2<\phi_i/f_a<\pi$, respectively.
If $4M_{\rm pl}^2V_{,\phi \phi}(\phi_i)/(3V(\phi_i))>1$ 
(i.e., $K^2<0$), we can derive the similar expression of $w$
by setting $K=i\hat{K}$ and $\hat{K}=
\sqrt{4M_{\rm pl}^2V_{,\phi \phi}(\phi_i)/(3V(\phi_i))-1}$ \cite{Chiba09}.
For a phantom field and a thawing k-essence field,
analytic solutions of $w$ similar to (\ref{waap}) were derived 
in Refs.~\cite{Dutta2,SchChiba}.

In Fig.~\ref{fig3} we plot numerically integrated solutions of 
$w$ as well as analytic solutions based on (\ref{waap})
for three different values of $K$.
As long as $K \lesssim 10$ and $w_0 \lesssim -0.3$, 
the analytic estimation of (\ref{waap}) is sufficiently trustable.
For larger $K$, the field mass squared $|m_{\phi}^2|$ increases
and hence the variation of $w$ around today is more significant.
For the validity of the Taylor expansion used to derived
the analytic solution (\ref{waap}),
we require the condition $|K-1|<{\cal O}(1)$.
For the models with $V_{,\phi \phi}>0$, the analytic
solutions start to lose the accuracy for $K$ smaller than 0.5.

The observational constraints on thawing models
have been carried out in Refs.~\cite{Dutta,SchChiba,Clemson,Gupta,CDT}.
If the three parameters $w_0$, $\Omega_{\phi 0}$, and $K$ are 
varied in the likelihood analysis with the prior $0.1<K<10$, 
the constraints on $K$ are generally weak.
After the marginalization over $K$ without any prior 
on $w_0$, Chiba {\it et al.} \cite{CDT} derived the bounds
$-2.18<w_0<-0.893$ and $0.703<\Omega_{\phi 0}<0.735$ (95\,\% CL).
If we put the prior $w_0>-1$, the field equation of state
is constrained to be $w_0<-0.849$ (68~\% CL) and 
$w_0<-0.695$ (95~\% CL). 
Although there is no statistical evidence that the models with 
$w_0>-1$ are favored over the $\Lambda$CDM, 
the thawing models with $-1<w_0<-0.7$
are not yet ruled out observationally.

\section{Constraints from the large-scale structure}
\label{aba:sec4}

In addition to the background observational constraints discussed in 
Sec.\,\ref{aba:sec3}, the quintessence models can be distinguished from 
the $\Lambda$CDM by considering the evolution of cosmological 
perturbations. The peculiar velocity of inward collapse motion 
of the large-scale structure is directly related to the growth rate of 
the matter density contrast $\delta_m$. Then the measurement from redshift-space distortions (RSD) 
of clustering pattern of galaxies can constrain the growth history 
of the large-scale structure.
The galaxy redshift surveys provide bounds on the 
growth rate $f(z)$ or $f(z) \sigma_8 (z)$ in terms of the
redshift $z=1/a-1$, where $f=d \ln \delta_m/d \ln a$
and $\sigma_8$ is the rms amplitude of $\delta_m$
at the comoving scale $8\,h^{-1}$ Mpc 
($h$ is the normalized Hubble constant 
$H_0=100\,h$\,km\,sec$^{-1}$\,Mpc$^{-1}$).
It is then convenient to derive analytic solutions of $f(z)$
and $f(z) \sigma_8 (z)$ to confront quintessence models with 
the observations of RSD.

Let us consider scalar metric perturbations $\Phi$ about the flat FLRW 
background. Neglecting the anisotropic stress, the metric in 
the Newtonian gauge is given by \cite{Bardeen}
\be
ds^{2}=-(1-2\Phi)\, dt^{2}+a^{2}(t)(1+2\Phi) d {\bm x}^2\,.
\label{permet}
\ee
We decompose the matter density $\rho_m$ 
into the background and inhomogeneous parts, 
as $\rho_m=\rho_m (t)+\delta \rho_m (t, {\bm x})$.
We also define 
\be
\delta_m \equiv \delta \rho_m/\rho_m\,,\qquad
\theta_m \equiv \nabla^2 v_m/(aH)\,,
\ee
where $v_m$ is the rotational-free velocity potential 
of non-relativistic matter.
In the Fourier space the matter perturbations 
satisfy \cite{Bardeen,Kodama}
\ba
& & \delta_m'=-3 \Phi'-\theta_m\,,
\label{dm1}\\
& & \theta_m'+\left( 2+\frac{H'}{H} \right)\theta_m=
-\left( \frac{k}{aH} \right)^2 \Phi\,,
\label{dm2}
\ea
where $k$ is a comoving wave number and a prime 
represents a derivative with respect to $N=\ln a$.
Taking the $N$-derivative of Eq.~(\ref{dm1}) 
and using Eq.~(\ref{dm2}), we obtain
\begin{equation}
\delta_m''+\left( 2+\frac{H'}{H} \right) \delta_m'
-\left( \frac{k}{aH} \right)^2 \Phi
=-3 \left[ \Phi''+\left( 2+\frac{H'}{H} \right)\Phi' \right]\,.
\label{delmeq0}
\end{equation}

Provided that quintessence does not cluster, we can neglect 
the contribution of quintessence perturbations relative to 
matter perturbations. In this case the Poisson equation 
is approximately given by 
\be
\left( \frac{k}{aH} \right)^2 \Phi \simeq 
\frac{3}{2} \Omega_m \hat{\delta}_m\,,
\label{Poisson}
\ee
where $\hat{\delta}_m \equiv \delta_m+3(aH/k)^2 \theta_m$
is the rest-frame gauge-invariant density perturbation.
For the modes deep inside the Hubble radius ($k \gg aH$) 
relevant to the large-scale structure, the r.h.s. of Eq.~(\ref{delmeq0}) 
can be neglected relative to the l.h.s. of it.
Using the approximate relation $\hat{\delta}_m \simeq \delta_m$
together with Eqs.~(\ref{dothquin}) and (\ref{Poisson}), 
we find that Eq.~(\ref{delmeq0}) reduces to 
\be
\delta_m''+\frac12 \left(1-3w \Omega_{\phi} \right) \delta_m'
-\frac32 \Omega_m \delta_m \simeq 0\,.
\label{delmeq}
\ee

In the following we derive analytic solutions for the growth rate 
$f$ and $f\sigma_8$ as functions of the 
redshift $z$. Since $\Omega_{\phi} \ll 1$ at the 
early cosmological epoch,  
we expand the quintessence equation of state 
in terms of $\Omega_{\phi}$, as
\be
w=w_0+\sum_{n=1}^{\infty} w_n (\Omega_{\phi})^n\,.
\label{wexpand}
\ee
We introduce the growth index $\gamma$, as 
$f=\delta_m'/\delta_m=(\Omega_m)^{\gamma}
=(1-\Omega_{\phi})^{\gamma}$ \cite{Peebles,Wang98,Linder}.
Using Eq.~(\ref{quinOme}) with $w_m=0$,
Eq.~(\ref{delmeq}) reads \cite{Wang98}
\begin{equation}
\hspace{-2cm}
3w \Omega_{\phi} (1-\Omega_{\phi}) 
\ln  (1-\Omega_{\phi})
\frac{d \gamma}{d \Omega_{\phi}} 
=\frac12-\frac32 w(1-2\gamma)\Omega_{\phi}
+(1-\Omega_{\phi})^{\gamma}-\frac32 
(1-\Omega_{\phi})^{1-\gamma}\,.
\label{gamOme}
\end{equation}

The solution of Eq.~(\ref{gamOme}) can be derived
by expanding $\gamma$ in terms of $\Omega_{\phi}$, as
$\gamma=\gamma_0+\sum_{n=1}^{\infty} \gamma_n (\Omega_{\phi})^n$.
On using the expansion (\ref{wexpand}) as well, we obtain \cite{Gong,Jailson}
\begin{equation}
\gamma=\frac{3(1-w_0)}{5-6w_0}+\frac32 
\frac{(1-w_0)(2-3w_0)+2w_1(5-6w_0)}{(5-6w_0)^2 (5-12w_0)}
\Omega_{\phi}+{\cal O}(\Omega_{\phi}^2)\,.
\label{gammaap}
\end{equation}
If $w_0=-1$ and $w_1=0$, then we have 
$\gamma \simeq 0.545+7.29 \times 10^{-3} \Omega_{\phi}$.
Since the second term is much smaller than the first one, 
$\gamma$ is nearly constant.
Even for the models with $w_0=-1$ and $w_1=0.3$
(where the value of $w$ today is around $-0.8$), 
the variation of $\gamma$ is small: $\gamma \simeq 
0.545+1.21 \times 10^{-2} \Omega_{\phi}$.

The relation $\delta_m'/\delta_m=(1-\Omega_{\phi})^{\gamma}$
can be written in the form
\begin{equation}
\frac{d}{d \Omega_{\phi}} \ln \delta_m
=-\frac{(1-\Omega_{\phi})^{\gamma-1}}
{3w \Omega_{\phi}}\,.
\label{delmeqa}
\end{equation}
Under the approximation that $\gamma$ is constant, 
the term $(1-\Omega_{\phi})^{\gamma-1}$ can be expanded
around $\Omega_{\phi}=0$, as 
\begin{equation}
(1-\Omega_{\phi})^{\gamma-1}
=1+\sum_{n=1}^{\infty} \alpha_n
(\Omega_{\phi})^n\,,\qquad 
\alpha_n=\frac{(-1)^n}{n!} \prod_{i=1}^{n}
(\gamma-i)\,.
\end{equation}
We also expand $1/w$ in the form 
$1/w=(1/w_0) [1+\sum_{n=1}^{\infty} \beta_{n} (\Omega_{\phi})^n]$, 
where $\beta_n$ can be expressed in terms of $w_i$ ($i=0,1,2,\cdots$).
Then Eq.~(\ref{delmeqa}) is written as
\begin{equation}
\hspace{-1cm}
\frac{d}{d \Omega_{\phi}} \ln \delta_m
=-\frac{1}{3w_0 \Omega_{\phi}}
\left[ 1+\sum_{n=1}^{\infty} c_n(\Omega_{\phi})^n \right]\,,
\qquad
c_n=\sum_{i=0}^n \alpha_{n-i} \beta_{i}\,,
\label{delmeqa2}
\end{equation}
with $\alpha_0=\beta_0=1$.
Integration of Eq.~(\ref{delmeqa2}) leads to the 
following solution 
\begin{equation}
\delta_m=\delta_{m0} \exp \left\{ \frac{1}{3w_0}
\left[ \ln \frac{\Omega_{\phi 0}}{\Omega_{\phi}}+\sum_{n=1}^{\infty}
\frac{c_n}{n} \left( (\Omega_{\phi 0})^n- (\Omega_{\phi})^n \right) \right]
\right\}\,,
\label{delmso}
\end{equation}
where $\delta_{m0}$ is the today's value of $\delta_m$. 

The perturbation $\delta_g$ of galaxies is related to $\delta_m$, 
as $\delta_g=b \delta_m$, where $b$ is a bias factor.
The galaxy power spectrum ${\cal P}_g ({\bm k})$ in 
the redshift space can be expressed 
as \cite{Kaiser,Tegmark06}
\begin{equation}
{\cal P}_g ({\bm k})={\cal P}_{gg} ({\bm k})
-2\mu^2 {\cal P}_{g \theta} ({\bm k})
+\mu^4 {\cal P}_{\theta \theta} ({\bm k})\,,
\label{Pred}
\end{equation}
where $\mu={\bm k} \cdot {\bm r}/(kr)$ is the cosine 
of the angle of the momentum vector ${\bm k}$ 
to the line of sight (vector ${\bm r}$),
${\cal P}_{gg} ({\bm k})$ and ${\cal P}_{\theta \theta} ({\bm k})$ 
are the real space power spectra 
of galaxies and $\theta$, respectively, 
and ${\cal P}_{g \theta} ({\bm k})$ is the cross 
power spectrum of galaxy-$\theta$ fluctuations
in real space.
Neglecting the variation of $\Phi$ in Eq.~(\ref{dm1}), 
it follows that 
\begin{equation}
\theta_m \simeq -f \delta_m\,.
\label{continuity}
\end{equation}
The three spectra ${\cal P}_{gg}$, ${\cal P}_{g \theta}$, 
and ${\cal P}_{\theta \theta}$ depend on $(b \delta_m)^2$, 
$(b \delta_m)(f \delta_m)$, and $(f \delta_m)^2$, respectively.
Provided that the growth of perturbations is scale-independent, 
the constraints on $b \delta_m$ and $f \delta_m$ at some scale 
translate into those on $b \sigma_8$ and $f \sigma_8$.
The quantity $f \sigma_8$ is useful because 
it does not include the bias factor $b$.

Normalizing $\delta_{m0}$ in terms of $\sigma_8 (z=0)$ in 
Eq.~(\ref{delmso}), we obtain \cite{Jailson}
\begin{equation}
\hspace{-2cm}
f(z) \sigma_8 (z)=
(1-\Omega_{\phi})^{\gamma}\,\sigma_8 (z=0)\, \exp \left\{ \frac{1}{3w_0}
\left[ \ln \frac{\Omega_{\phi 0}}{\Omega_{\phi}}+\sum_{n=1}^{\infty}
\frac{c_n}{n} \left( (\Omega_{\phi 0})^n- (\Omega_{\phi})^n \right) \right]
\right\}\,.
\label{fsigma8}
\end{equation}
The 0-th order solution of $\Omega_{\phi}$ corresponds to 
$w=w_0$, in which case
$\Omega_{\phi}^{(0)}=\Omega_{\phi 0} (1+z)^{3w_0}/
[1-\Omega_{\phi 0}+\Omega_{\phi 0} (1+z)^{3w_0}]$.
Using the iterative solution $w=w_0+w_1 \Omega_{\phi}^{(0)}$, 
we obtain the first-order solution 
\begin{equation}
\Omega_{\phi}^{(1)}=\frac{\Omega_{\phi 0} (1+z)^{3w_0}
[1-\Omega_{\phi 0}+\Omega_{\phi 0} (1+z)^{3w_0}]^{w_1/w_0}}
{1-\Omega_{\phi 0}+\Omega_{\phi 0} (1+z)^{3w_0}
[1-\Omega_{\phi 0}+\Omega_{\phi 0} (1+z)^{3w_0}]^{w_1/w_0}}\,.
\label{Omegax1}
\end{equation}
We can continue the similar iterative processes, 
but it is practically sufficient to exploit
the 1-st order solution (\ref{Omegax1}) for the evaluation 
of $\Omega_\phi$ in Eq.~(\ref{fsigma8}).

There exists a quintessence potential in which 
$w$ is constant \cite{Saini,Jailson} 
(see also Ref.~\cite{darkreview1}).
In this case we have 3 free parameters 
$w_0$, $\Omega_{\phi 0}$, $\sigma_8 (z=0)$ 
in the expression of $f(z) \sigma_8(z)$. 
In tracking quintessence models the coefficients 
$w_n$ ($n \geq 1$) are expressed in terms of 
$w_0=w_{(0)}$, so there are also 3 free parameters 
$w_0$, $\Omega_{\phi 0}$, and $\sigma_8(z=0)$. 
In these cases it was shown that the analytic result 
(\ref{fsigma8}) up to 7-th order terms of $c_n$ is sufficiently accurate to reproduce 
full numerical solutions in high precision \cite{Jailson}.
In thawing quintessence models, when the variation of $w$ 
is fast at late times, the analytic solution (\ref{fsigma8}) is 
not very accurate unless higher-order terms of 
$c_n$ are taken into account.

In constant $w$ models the observational data of RSD up to 2012 
place the bound $-1.245<w<-0.347$ (68\,\%\,CL),
whereas in the tracking models the tracker equation of state 
is constrained to be $-1.288<w_{(0)}<-0.214$  (68\,\%\,CL).
Although these constraints are still weak, this situation
will be improved in future high-precision measurements.

\section{Particle physics models of quintessence}
\label{aba:sec5}

There have been many attempts to construct particle physics 
model of quintessence in the framework of supersymmetric theories.
Binetruy \cite{Binetruy} showed that the inverse power-law potential 
(\ref{inversepo}) appears in a globally supersymmetric $SU (N_c)$ 
gauge theory with $N_c$ colors and the condensation of $N_f$ flavors.
In this theory the power $p$ in Eq.~(\ref{inversepo}) is given by 
$p=2(N_c+N_f)/(N_c-N_f)$, which is larger than 2 under the 
condition $N_c \geq N_f>0$.
Since $p$ is constrained to be smaller than $0.075$ \cite{CDT}, 
this scenario is not compatible with the current observational data.

In the presence of gravity, any globally supersymmetric
theory reduces to a locally supersymmetric supergravity theory. 
In supergravity the four-dimensional effective action 
is given by \cite{sugra}
\begin{equation}
S=\int d^{4}x\sqrt{-g}\left[\frac{M_{\rm pl}^2}{2}R
-K_{ij^{*}}\partial_{\mu} \vp^i\partial^{\mu}\vp^{j^{*}}
-V(\vp,\vp^{*})
\right]\,,
\label{sugraaction}
\end{equation}
where $\vp$ are chiral scalar fields, and 
$K^{ij^{*}}$ is an inverse of the derivative of the so-called
K\"{a}hler potential $K$, i.e., 
$K_{ij^{*}} \equiv \partial^{2}K/\partial\vp^{i}\partial\vp^{j^{*}}$.
The effective cosmological constant $V$ is expressed 
in terms of $K$ and the superpotential $W$, as
\begin{equation}
V(\vp,\vp^{*})=e^{K/M_{\rm pl}^2}
\left[D_{i}W(K^{ij^{*}})(D_{j}W)^{*}
-3|W|^{2}/M_{\rm pl}^2\right]\,,
\label{superpo}
\end{equation}
where $D_{i}W \equiv \partial W/\partial \vp^{i}
+(W/M_{\rm pl}^2) \partial K/\partial \vp^{i}$.

The last term in Eq.~(\ref{superpo}) is negative and hence this can be 
an obstacle to realize a positive vacuum energy required for dark energy.
For example, Brax and Martin \cite{Brax99} chose a superpotential  
$W=\Lambda^{3+\alpha}\vp^{-\alpha}$ 
(motivated by the fermion condensate gauge theory 
mentioned above) 
and a flat K\"{a}hler potential $K=\vp\vp^{*}$, 
but in this case the potential $V$ becomes 
negative for $\phi \sim M_{\rm pl}$.
This problem can be avoided by imposing 
$\langle W \rangle=0$ \cite{Brax99}, 
but such a constraint is generally difficult to be compatible with 
the models of supersymmetry breaking.
The K\"{a}hler potential of the form
$K=-M_{\rm pl}^2 \ln [\left( \vp+\vp^{*} \right)/M_{\rm pl}]$,
which is present at tree level for both the dilaton and moduli 
fields in string theory, 
can allow the possibility of canceling the negative term 
$(-3|W|^{2}/M_{\rm pl}^2)$.
Introducing a new field $\phi=(M_{\rm pl}/\sqrt{2})\,\ln( \vp/M_{\rm pl})$ 
in this case, the kinetic term in the action (\ref{sugraaction}) 
reduces to the canonical form 
${\cal L}_{\rm kin}=-\partial^{\mu} \phi \partial_{\mu} \phi/2$.
For the choice $W=\Lambda^{3+\alpha}\vp^{-\alpha}$, the potential 
(\ref{superpo}) reads \cite{Rosati}
\begin{equation}
V(\phi)=M^{4} e^{-\sqrt{2}\beta\,\phi/M_{\rm pl}}\,,
\end{equation}
where $\beta\equiv2\alpha+1$ and 
$M^{4} \equiv M_{\rm pl}^{-\beta-1}\Lambda^{\beta+5}(\beta^{2}-3)/2$. 
The positivity of the potential requires the condition $\beta>\sqrt{3}$.
Then the slope of the exponential potential, $\lambda\equiv\sqrt{2}\beta$, 
satisfies the condition $\lambda>\sqrt{6}$.
In this case there exists a scaling solution along which 
$\Omega_{\phi}=3(1+w_m)/\lambda^2$ is constant 
with $w=w_m$, but the potential needs to be 
modified at late times to realize the cosmic acceleration.

Copeland {\it et al.} \cite{Rosati} tried to construct a viable quintessence 
potential by choosing
\begin{equation}
K=M_{\rm pl}^2 \left[ \ln \left( \vp+\vp^{*}\right)/M_{\rm pl} \right]^{2}\,,
\qquad
W=\Lambda^{3+\alpha}\vp^{-\alpha}\,,
\end{equation}
where the field $\vp$ is assumed to be real. 
The kinetic term becomes canonical by introducing a new scalar
field, $\phi \equiv \int\sqrt{2K_{\vp\vp^{*}}}\,\rd\vp=
-(2M_{\rm pl}/3) \left[1-\ln(2 \vp/M_{\rm pl}) \right]^{3/2}$. 
The field potential is given by 
\begin{equation}
V=m^4 \left[2Y^{2}+(4\alpha-7)Y+2(\alpha-1)^{2}\right]
e^{(1-Y)^{2}-2\alpha(1-Y)}/Y\,,
\label{Vsugra}
\end{equation}
where $m^4 \equiv 2^{2\alpha}M_{\rm pl}^{-2-2\alpha} 
\Lambda^{6+2\alpha}$ and 
\begin{equation}
Y \equiv 1-\ln( 2\vp/M_{\rm pl})=[-(3/2)(\phi/M_{\rm pl})]^{2/3}\,.
\end{equation}
The field exists in the region $-\infty<\phi<0$, which corresponds 
to $0<Y<\infty$. For $|\phi| \ll M_{\rm pl}$ and 
$|\phi| \gg M_{\rm pl}$, the potential behaves as 
$V\propto(-\phi)^{-2/3}$ and 
$V\propto(-\phi)^{2/3}e^{(-\phi/M_{\rm pl})^{4/3}}$ respectively.
In the intermediate region there exists a potential minimum 
with a positive energy density.
For the initial conditions satisfying $|\phi| \ll M_{\rm pl}$
the quintessence potential is approximately given by 
$V(\phi) \propto (-\phi)^{-2/3}$ in the early cosmological epoch, 
so that the field exhibits a tracking behavior.
If the field is initially in the region $|\phi| \gg M_{\rm pl}$, the 
contribution of the exponential potential
is important. In this case a scaling-like behavior can be 
realized during the radiation and matter eras \cite{Rosati}. 
As the field approaches the potential minimum, the Universe 
enters the epoch of cosmic acceleration.

A general problem for supersymmetric quintessence models is that 
supersymmetry must be broken if it is to be realized at all in nature. 
In the gravity and gauge mediated scenarios, 
the supersymmetry breaking is supposed to occur 
for the energy scale larger than $\langle F \rangle^{1/2} \gtrsim10^{10}$\,GeV
and $\langle F \rangle^{1/2} \gtrsim10^{4}$\,GeV (where $F^2$
is the first term in Eq.~(\ref{superpo}), i.e., 
$F^2 \equiv e^{K/M_{\rm pl}^2} D_{i}W(K^{ij^{*}})(D_{j}W)^{*}$), respectively,
to lift the masses of supersymmetric scalar particles above $10^{2}$~GeV. 
In order to give a negligible vacuum energy in Eq.~(\ref{superpo}) 
we require that the superpotential takes the form 
$W \sim \langle F \rangle M_{\rm pl} \sim m_{3/2}M_{\rm pl}^2$, 
where $m_{3/2}$ is the gravitino mass \cite{Rosati}.
Then the superpotential $W=\Lambda^{3+\alpha} \vp^{-\alpha}$ used above  
gets corrected by the term $m_{3/2}M_{\rm pl}^2$. 
This gives rise to the correction of the order of $m_{3/2}^{2}M_{\rm pl}^2$
to the quintessence potential, so that the flatness of the potential 
required for the late-time cosmic acceleration can be spoiled.

Although this problem looks serious, unconventional
supersymmetry breaking models in string theory may overcome 
this problem. In Ref.~\cite{Witten} it was suggested that we
may live in a four-dimensional world with unbroken supersymmetry. 
In this scenario the mass splitting between the superpartners occurs
as a result of the excitations of the system while maintaining a supersymmetric
ground state. Then we do not need to worry about the
contribution of the supersymmetry breaking terms to 
the potential. 

There are also some supergravity models in which the above mentioned 
problem can be avoided. In the framework of extended supergravity 
models \cite{Kallosh,Fre} 
the mass squared of any light scalar fields can be quantized in unit
of squared of the Hubble constant $H_{0}$ of de Sitter solutions. 
The de Sitter solutions correspond to the extrema of
an effective potential $V(\phi)$ of a scalar field $\phi$. 
Around the extremum at $\phi=0$ the field potential is given by 
$V(\phi)=\Lambda+(1/2)m_{\phi}^{2}\phi^{2}$
with $\Lambda>0$. In extended supergravity theories the mass $m_{\phi}$
is related to $\Lambda$ via the relation 
$m_{\phi}^{2}=n\Lambda/(3M_{{\rm pl}}^{2})$,
where $n$ is an integer.
Since $H_{0}^{2}=\Lambda/(3M_{{\rm pl}}^{2})$ for de Sitter solutions, 
$m_{\phi}^{2}=nH_{0}^{2}$. 
In the ${\cal N}=2$ and ${\cal N}=8$ extended supergravity theories we have 
$n=6$ and $n=-6$ respectively \cite{Fre,Kallosh}, so that the field potentials are
\begin{equation}
V(\phi)=3H_{0}^{2}M_{{\rm pl}}^{2}\left[1 \pm 
\left(\phi/M_{{\rm pl}}\right)^{2}\right]\,.
\label{Vphide}
\end{equation}
The energy scale of the supersymmetry
breaking is determined by the constant $\Lambda$.  
If the potential (\ref{Vphide}) is responsible for dark energy, 
we require that 
$\Lambda \approx H_{0}^{2}M_{{\rm pl}}^{2} \approx 10^{-47}\,{\rm GeV}^{4}$.
The supersymmetry breaking scale is so small that the 
ultra light mass of the order of $10^{-33}$\,eV can be protected
against quantum corrections. 

The PNGB models based on the potential (\ref{pngbpo}) also allow to 
protect the light mass of quintessence by the $U(1)$ symmetry. 
An example of a very light PNGB is the so-called axion field, which was
originally introduced to address the strong CP problem \cite{PQ}.
When a global $U(1)$ symmetry is spontaneously broken,
the axion appears as an angular field $\phi$ with an expectation
value $\langle\vp\rangle=f_{a}e^{i\phi/f_{a}}$ of a complex scalar
at a scale $f_{a}$. 
In string theory there are many light axions, possibly
populating each decade of mass down to the scale 
$H_0 \approx 10^{-33}$~eV \cite{Dimo}. 
In the limit $\mu \to 0$ the potential vanishes, 
so that the symmetry becomes exact.
The radiative corrections to $V$ do not give rise to an explicit 
symmetry breaking term because they are proportional
to $\mu^{4}$. Hence the small mass associated with dark energy 
can be protected against radiative corrections.

If the PNGB potential (\ref{pngbpo}) is responsible for the cosmic
acceleration today, we require that $H_{0}^{2}\approx\mu^{4}/M_{{\rm pl}}^{2}$ 
and hence $\mu \approx 10^{-3}$\,eV.
The field mass squared around $\phi=0$ can be estimated as
$m_{\phi}^{2} \approx -(M_{{\rm pl}}^{2}/f_a^{2})H_{0}^{2}$.
The slow-roll condition, $|M_{\rm pl}^2 V_{,\phi\phi}/V| \lesssim 1$,
translates into $f_a \gtrsim M_{{\rm pl}}$. Then the field mass is constrained 
to be $|m_{\phi}| \lesssim H_{0}$, so that the field starts to evolve only 
recently. As we studied in Sec.\,\ref{thawingsec}, this belongs to the class of 
thawing quintessence models.

In supersymmetric theories there have been a number of attempts to explain 
the small energy scale $\mu \approx 10^{-3}$\,eV \cite{Nomura,Choi,Kim,Hall}. 
Hall {\it et al.} \cite{Hall} tried to relate $\mu$ with
two fundamental scales, the Planck scale $M_{{\rm pl}}\approx10^{18}$\,GeV
and the electroweak scale $v\approx10^{3}$\,GeV. There is the induced
seesaw scale $v^{2}/M_{{\rm pl}}\approx10^{-3}$\,eV, 
which is of the same order of $\mu$. 
If we assume the relation $\mu \approx v^{2}/M_{{\rm pl}}$ and $f_a=M_{\rm pl}$, 
it follows that  $|m_{\phi}^{2}| \approx \mu^{4}/f_a^{2} \approx v^{8}/M_{{\rm pl}}^{6}$.
This gives rise to the mass of the order
$|m_{\phi}|\approx v^{4}/M_{{\rm pl}}^{3} \approx 10^{-33}~{\rm eV}$.

In order to justify the relation $\mu \approx v^{2}/M_{{\rm pl}}$, Hall {\it et al.} \cite{Hall}
proposed supersymmetric models with an axion in a hidden sector.
In this set up the axion $\phi$ has interactions with the quarks $q$, $q^c$
in the form ${\cal L}_{\rm int}=m_q q q^c e^{i\phi/f_a}$ at a scale $M$, where $m_q$ is 
the quark mass of the order of the effective supersymmetry breaking scale 
$m_B=v^2/M_{\rm pl}$. If at least one of the quark flavors 
has a mass smaller than the order of $M$, a quark condensate
forms such that $\langle qq^{c}\rangle\approx M^{3}e^{i\tilde{\phi}/M}$
with an angular field $\tilde{\phi}$.
This gives rise to the axion potential 
$V=m_q M^3 \cos (\phi/f_a+\tilde{\phi}/M)$, where $M$ is close to $m_B$.
Then the scale $\mu$ is of the order of $\mu \approx m_B=v^2/M_{\rm pl}$.

In summary the thawing models based on the potentials (\ref{pngbpo}) and (\ref{Vphide})
are good candidates of quintessence from the theoretical point of view.
 
\section{Conclusions}
\label{aba:sec6}

We have reviewed theoretical and observational aspects 
of quintessence. We classified quintessence models in terms of
the evolution of the field equation of state $w$.

In tracking models the solutions with different initial conditions converge 
to a common trajectory characterized by the analytic solution (\ref{wtracker}). 
A typical example of this class is the  potential 
$V(\phi)=M^{4+p}\phi^{-p}$ ($p>0$), in which case 
$w$ is nearly constant 
($w_{(0)}=-2/(p+2)$) during the matter era.
The joint data analysis of SN Ia, CMB, and BAO gives the bound
$w_{(0)}<-0.964$ (95\,\%\,CL) and hence the deviation from 
the $\Lambda$CDM is small.
The inverse power-law potential appears in a fermion condensate model of 
a globally supersymmetric gauge theory, but the theoretical values of 
$p$ are larger than those constrained by observations.

The exponential potential $V(\phi)=V_0 e^{-\lambda \phi/M_{\rm pl}}$
gives rise to a scaling solution along which $w=w_m$ and 
$\Omega_{\phi}=3(1+w_m)/\lambda^2$. 
Under the condition $\lambda^2>3(1+w_m)$
the scaling solution is an attractor during the radiation 
and matter eras, but it does not 
exit to the epoch of cosmic acceleration. 
This problem can be alleviated for the double exponential 
potential (\ref{doublepo}) or for the potential (\ref{Vsugra}) 
appearing in the context of supergravity. The likelihood analysis for 
the potential (\ref{doublepo}) with $\lambda_2=0$ shows 
that the transition from $w=0$ to $w=-1$ needs to 
occur at the early cosmological epoch ($a_t<0.23$ (95\,\%\,CL) 
according to the parametrization (\ref{LHpara}) with 
$w_p=0$ and $w_f=-1$).

In thawing models there is an analytic solution (\ref{waap}) 
of $w$ written in terms of the three parameters $w_0$, 
$\Omega_{\phi 0}$, and $K$.
The parameter $K$ is related to the mass of quintessence.
We require the condition $K \lesssim 10$ to avoid the 
rapid roll down of the field along the potential.
Under the prior $w_0>-1$, the today's field equation of state
is constrained to be $w_0<-0.695$ (95\,\%\,CL) from the 
joint data analysis of SN Ia, CMB, and BAO.
The potential (\ref{pngbpo}) of PNGB and the potentials (\ref{Vphide}) appearing 
in extended supergravity theories belong to the class of thawing 
models. In these models,  the small field mass $m_{\phi}$ associated
with dark energy can be protected against radiative corrections
due to underlying symmetries.

In order to confront quintessence models with the observations
of redshift-space distortions of clustering pattern of galaxies,
we derived analytic formulas for the growth rate $f(z)$ as well as 
$f(z) \sigma_8(z)$ of matter density perturbations.
These are useful to place constraints on the quintessence 
equation of state. We expect that future high-precision
observations of RSD, combined with other measurements 
such as SN Ia, CMB, BAO, and weak lensing, will allow us to 
distinguish quintessence from $\Lambda$CDM.

\section*{Acknowledgments}

I thank David Langlois to invite me to write this article 
for a special issue of Classical and Quantum Gravity on 
scalars and gravity.
I am also grateful to Takeshi Chiba and Antonio De Felice
for useful discussions.



\section*{References}
\begin{thebibliography}{10}


\bibitem{Riess}
A.~G.~Riess {\it et al.},
Astron.\ J.\  {\bf 116}, 1009 (1998).

\bibitem{Perlmutter}
S.~Perlmutter {\it et al.},
Astrophys.\ J.\  {\bf 517}, 565 (1999).

\bibitem{WMAP1} 
D.~N.~Spergel {\it et al.}  [WMAP Collaboration],
Astrophys.\ J.\ Suppl.\  {\bf 148}, 175 (2003).
 
\bibitem{Planck} 
P.~A.~R.~Ade {\it et al.}  [Planck Collaboration],
arXiv:1303.5076 [astro-ph.CO].
 
\bibitem{BAO1}
D.~J.~Eisenstein {\it et al.}  [SDSS Collaboration],
Astrophys.\ J.\  {\bf 633}, 560 (2005).

\bibitem{Weinberg}
S.~Weinberg,
Rev.\ Mod.\ Phys.\  {\bf 61}, 1 (1989).

\bibitem{KKLT} 
S.~Kachru, R.~Kallosh, A.~D.~Linde and S.~P.~Trivedi,
Phys.\ Rev.\ D {\bf 68}, 046005 (2003).

\bibitem{darkreview1}
V.~Sahni and A.~A.~Starobinsky,
Int.\ J.\ Mod.\ Phys.\  D {\bf 9}, 373 (2000).

\bibitem{darkreview2}
S.~M.~Carroll,
Living Rev.\ Rel.\  {\bf 4}, 1 (2001);
P.~J.~E.~Peebles and B.~Ratra,
Rev.\ Mod.\ Phys.\  {\bf 75}, 559 (2003);
T.~Padmanabhan,
Phys.\ Rept.\  {\bf 380}, 235 (2003); 
E.~J.~Copeland, M.~Sami and S.~Tsujikawa,
Int.\ J.\ Mod.\ Phys.\ D {\bf 15}, 1753 (2006);
S.~Tsujikawa,
arXiv:1004.1493 [astro-ph.CO];
L.~Amendola and S.~Tsujikawa,
{\it Dark energy: Theory and Observations},
Cambridge University Press (2010).

\bibitem{Fujii}
Y.~Fujii, Phys.\ Rev.\ D {\bf 26}, 2580 (1982);
L.~H.~Ford,
Phys.\ Rev.\ D {\bf 35}, 2339 (1987);
C.~Wetterich, Nucl.\ Phys \ B. {\bf 302}, 668 (1988).

\bibitem{Ratra}
B.~Ratra and P.~J.~E.~Peebles,
Phys.\ Rev.\ D {\bf 37}, 3406 (1988).

\bibitem{CSN}
T.~Chiba, N.~Sugiyama and T.~Nakamura,
Mon.\ Not.\ Roy.\ Astron.\ Soc.\  {\bf 289}, L5 (1997).

\bibitem{Ferreira}
P.~G.~Ferreira and M.~Joyce,
Phys.\ Rev.\ Lett.\  {\bf 79}, 4740 (1997);
Phys.\ Rev.\ D {\bf 58}, 023503 (1998).

\bibitem{CLW} 
E.~J.~Copeland, A.~R.~Liddle and D.~Wands,
Phys.\ Rev.\ D {\bf 57}, 4686 (1998).

\bibitem{Caldwell}
R.~R.~Caldwell, R.~Dave and P.~J.~Steinhardt,
Phys.\ Rev.\ Lett.\  {\bf 80}, 1582 (1998).

\bibitem{Zlatev} 
I.~Zlatev, L.~M.~Wang and P.~J.~Steinhardt,
Phys.\ Rev.\ Lett.\  {\bf 82}, 896 (1999).

\bibitem{kes1}
T.~Chiba, T.~Okabe and M.~Yamaguchi,
Phys.\ Rev.\  D {\bf 62}, 023511 (2000).

\bibitem{kes2}
C.~Armendariz-Picon, V.~F.~Mukhanov and P.~J.~Steinhardt,
Phys.\ Rev.\ Lett.\  {\bf 85}, 4438 (2000);
Phys.\ Rev.\  D {\bf 63}, 103510 (2001).

\bibitem{chap}
A.~Y.~Kamenshchik, U.~Moschella and V.~Pasquier,
Phys.\ Lett.\  B {\bf 511}, 265 (2001).

\bibitem{moreview}
T.~P.~Sotiriou and V.~Faraoni,
Rev.\ Mod.\ Phys.\  {\bf 82}, 451 (2010);
A.~De Felice and S.~Tsujikawa,
Living Rev.\ Rel.\  {\bf 13}, 3 (2010);
S.~Tsujikawa,
Lect.\ Notes Phys.\  {\bf 800}, 99 (2010);
T.~Clifton, P.~G.~Ferreira, A.~Padilla and C.~Skordis,
Phys.\ Rept.\  {\bf 513}, 1 (2012).

\bibitem{Macorra}
A.~de la Macorra and G.~Piccinelli,
Phys.\ Rev.\  D {\bf 61}, 123503 (2000).

\bibitem{Nunes}
S.~C.~C.~Ng, N.~J.~Nunes and F.~Rosati,
Phys.\ Rev.\  D {\bf 64}, 083510 (2001).

\bibitem{Cora} 
P.~S.~Corasaniti and E.~J.~Copeland
Phys.\ Rev.\ D {\bf 67}, 063521 (2003).

\bibitem{CLinder} 
R.~R.~Caldwell and E.~V.~Linder,
Phys.\ Rev.\ Lett.\  {\bf 95}, 141301 (2005).

\bibitem{Linder06}
E.~V.~Linder,
Phys.\ Rev.\  D {\bf 73}, 063010 (2006).

\bibitem{SWZ}
P.~J.~Steinhardt, L.~M.~Wang and I.~Zlatev,
Phys.\ Rev.\ D {\bf 59}, 123504 (1999).

\bibitem{SchSen} 
R.~J.~Scherrer and A.~A.~Sen,
Phys.\ Rev.\ D {\bf 77}, 083515 (2008).

\bibitem{Dutta} 
S.~Dutta and R.~J.~Scherrer,
Phys.\ Rev.\ D {\bf 78}, 123525 (2008).

\bibitem{Chiba09} 
T.~Chiba,
Phys.\ Rev.\ D {\bf 79}, 083517 (2009).

\bibitem{Chiba10} 
T.~Chiba,
Phys.\ Rev.\ D {\bf 81}, 023515 (2010).

\bibitem{Kaiser} 
N.~Kaiser,
Mon.\ Not.\ Roy.\ Astron.\ Soc.\  {\bf 227}, 1 (1987).

\bibitem{Tegmark06} 
M.~Tegmark {\it et al.}  [SDSS Collaboration],
Phys.\ Rev.\ D {\bf 74}, 123507 (2006).

\bibitem{Wang98}
L.~-M.~Wang and P.~J.~Steinhardt,
Astrophys.\ J.\  {\bf 508}, 483 (1998).

\bibitem{Linder} 
E.~V.~Linder,
Phys.\ Rev.\ D {\bf 72}, 043529 (2005);
E.~V.~Linder and R.~N.~Cahn,
Astropart.\ Phys.\  {\bf 28}, 481 (2007).

\bibitem{Gong} 
Y.~Gong, M.~Ishak and A.~Wang,
Phys.\ Rev.\ D {\bf 80}, 023002 (2009).

\bibitem{Jailson} 
S.~Tsujikawa, A.~De Felice and J.~Alcaniz,
JCAP {\bf 1301}, 030 (2013).

\bibitem{Carrollqui}
S.~M.~Carroll,
Phys.\ Rev.\ Lett.\  {\bf 81}, 3067 (1998).

\bibitem{Kolda}
C.~F.~Kolda and D.~H.~Lyth,
Phys.\ Lett.\  B {\bf 458}, 197 (1999).

\bibitem{Frieman}
J.~A.~Frieman, C.~T.~Hill, A.~Stebbins and I.~Waga,
Phys.\ Rev.\ Lett.\  {\bf 75}, 2077 (1995).

\bibitem{Brax}
P.~Brax and J.~Martin,
Phys.\ Lett.\  B {\bf 468}, 40 (1999).

\bibitem{Nomura}
Y.~Nomura, T.~Watari and T.~Yanagida,
Phys.\ Lett.\  B {\bf 484}, 103 (2000).

\bibitem{Choi}
K.~Choi,
Phys.\ Rev.\  D {\bf 62}, 043509 (2000).

\bibitem{Kim}
J.~E.~Kim and H.~P.~Nilles,
Phys.\ Lett.\  B {\bf 553}, 1 (2003).

\bibitem{Hall}
L.~J.~Hall, Y.~Nomura and S.~J.~Oliver,
Phys.\ Rev.\ Lett.\  {\bf 95}, 141302 (2005).

\bibitem{CNR}
E.~J.~Copeland, N.~J.~Nunes and F.~Rosati,
Phys.\ Rev.\  D {\bf 62}, 123503 (2000).

\bibitem{Townsend}
P.~K.~Townsend,
JHEP {\bf 0111}, 042 (2001).

\bibitem{Panda}
S.~Panda, Y.~Sumitomo and S.~P.~Trivedi,
Phys.\ Rev.\ D {\bf 83}, 083506 (2011).


\bibitem{Lucchin}
F.~Lucchin and S.~Matarrese,
Phys.\ Rev.\ D {\bf 32}, 1316 (1985);
J.~J.~Halliwell,
Phys.\ Lett.\ B {\bf 185}, 341 (1987);
Y.~Kitada and K.~-i.~Maeda 
Phys.\ Rev.\ D {\bf 45}, 1416 (1992).


\bibitem{PWang} 
P.~-Y.~Wang, C.~-W.~Chen and P.~Chen
JCAP {\bf 1202}, 016 (2012).

\bibitem{CDT} 
T.~Chiba, A.~De Felice and S.~Tsujikawa,
Phys.\  Rev.\ D {\bf 87}, 083505 (2013).
 
\bibitem{Suzuki}
N.~Suzuki {\it et al.},
Astrophys.\ J.\  {\bf 746}, 85 (2012).

\bibitem{WMAP7} 
E.~Komatsu \textit{et al.} {[}WMAP Collaboration{]},
Astrophys.\ J.\ Suppl.\ \ {\bf 192}, 18 (2011).

\bibitem{SDSS7}
W.~J.~Percival {\it et al.},
Mon.\ Not.\ Roy.\ Astron.\ Soc.\ {\bf 401}, 2148 (2010).

\bibitem{BOSS} 
L.~Anderson {\it et al.},
Mon.\ Not.\ Roy.\ Astron.\ Soc.\  {\bf 428}, 1036 (2013).

\bibitem{BCN}
T.~Barreiro, E.~J.~Copeland and N.~J.~Nunes,
Phys.\ Rev.\  D {\bf 61}, 127301 (2000).

\bibitem{SahniWang}
V.~Sahni and L.~M.~Wang,
Phys.\ Rev.\  D {\bf 62}, 103517 (2000).

\bibitem{Albrecht}
A.~J.~Albrecht and C.~Skordis,
Phys.\ Rev.\ Lett.\  {\bf 84}, 2076 (2000).

\bibitem{Stewart}
S.~Dodelson, M.~Kaplinghat and E.~Stewart,
Phys.\ Rev.\ Lett.\  {\bf 85}, 5276 (2000).

\bibitem{Bean} 
R.~Bean, S.~H.~Hansen and A.~Melchiorri,
Phys.\ Rev.\ D {\bf 64}, 103508 (2001).

\bibitem{Linder05} 
E.~V.~Linder and D.~Huterer,
Phys.\ Rev.\ D {\bf 72}, 043509 (2005).

\bibitem{Bassett} 
B.~A.~Bassett, M.~Kunz, J.~Silk and C.~Ungarelli,
Mon.\ Not.\ Roy.\ Astron.\ Soc.\  {\bf 336}, 1217 (2002);
P.~S.~Corasaniti and E.~J.~Copeland,
Phys.\ Rev.\ D {\bf 67}, 063521 (2003);
B.~A.~Bassett, P.~S.~Corasaniti and M.~Kunz,
Astrophys.\ J.\  {\bf 617}, L1 (2004).

\bibitem{Piazza} 
F.~Piazza and S.~Tsujikawa,
JCAP {\bf 0407}, 004 (2004).

\bibitem{Waga} 
S.~Tsujikawa and M.~Sami, 
Phys.\ Lett.\ B {\bf 603}, 113 (2004);
L.~Amendola, M.~Quartin, S.~Tsujikawa and I.~Waga,
Phys.\ Rev.\ D {\bf 74}, 023525 (2006).

\bibitem{Tsuji06} 
S.~Tsujikawa,
Phys.\ Rev.\ D {\bf 73}, 103504 (2006).

\bibitem{assisted} 
A.~R.~Liddle, A.~Mazumdar and F.~E.~Schunck,
Phys.\ Rev.\ D {\bf 58}, 061301 (1998).

\bibitem{Junko} 
S.~A.~Kim, A.~R.~Liddle and S.~Tsujikawa,
Phys.\ Rev.\ D {\bf 72}, 043506 (2005);
G.~Calcagni and A.~R.~Liddle,
Phys.\ Rev.\ D {\bf 77}, 023522 (2008);
J.~Ohashi and S.~Tsujikawa,
Phys.\ Rev.\ D {\bf 80}, 103513 (2009).

\bibitem{Blais} 
D.~Blais and D.~Polarski,
Phys.\ Rev.\ D {\bf 70}, 084008 (2004).

\bibitem{Dutta2} 
S.~Dutta and R.~J.~Scherrer,
Phys.\ Lett.\ B {\bf 676}, 12 (2009).

\bibitem{SchChiba} 
T.~Chiba, S.~Dutta and  R.~J.~Scherrer,
Phys.\ Rev.\ D {\bf 80}, 043517 (2009).

\bibitem{Clemson} 
T.~Clemson and  A.~R.~Liddle,
Mon.\ Not.\ Roy.\ Astron.\ Soc.\  {\bf 395}, 1585 (2009).

\bibitem{Gupta} 
G.~Gupta, S.~Majumdar and A.~Sen,
Mon.\ Not.\ Roy.\ Astron.\ Soc.\  {\bf 420}, 1309 (2012).


\bibitem{Bardeen} 
J.~M.~Bardeen,
Phys.\ Rev.\ D {\bf 22}, 1882 (1980).

\bibitem{Kodama} 
H.~Kodama and M.~Sasaki,
Prog.\ Theor.\ Phys.\ Suppl.\  {\bf 78}, 1 (1984).

\bibitem{Peebles}
P.~J.~E.~Peebles, {\it Large-Scale Structure of the Universe,}
Princeton University Press (1980).

\bibitem{Saini} 
V.~Sahni, T.~D.~Saini, A.~A.~Starobinsky and U.~Alam,
JETP Lett.\  {\bf 77}, 201 (2003).


\bibitem{Binetruy} 
P.~Binetruy,
Phys.\ Rev.\ D {\bf 60}, 063502 (1999).

\bibitem{sugra} 
J.~Wess and J.~Bagger, 
{\it Supersymmetry and Supergravity}, 
Princeton University Press (1992).

\bibitem{Brax99} 
P.~Brax and J.~Martin
Phys.\ Lett.\ B {\bf 468}, 40 (1999).

\bibitem{Rosati} 
E.~J.~Copeland, N.~J.~Nunes and F.~Rosati,
Phys.\ Rev.\ D {\bf 62}, 123503 (2000).

\bibitem{Witten} 
E.~Witten,
hep-ph/0002297.

\bibitem{Kallosh} 
R.~Kallosh, A.~D.~Linde, S.~Prokushkin and M.~Shmakova,
Phys.\ Rev.\ D {\bf 65}, 105016 (2002);
Phys.\ Rev.\ D {\bf 66}, 123503 (2002).

\bibitem{Fre} 
P.~Fre, M.~Trigiante and A.~Van Proeyen,
Class.\ Quant.\ Grav.\  {\bf 19}, 4167 (2002).

\bibitem{PQ} 
R.~D.~Peccei and H.~R.~Quinn,
Phys.\ Rev.\ Lett.\  {\bf 38}, 1440 (1977).

\bibitem{Dimo} 
A.~Arvanitaki {\it et al.,}
Phys.\ Rev.\ D {\bf 81}, 123530 (2010).



\end{thebibliography}
\end{document}